\documentclass[aps,prb,superscriptaddress,10pt,twocolumn,nopacs,letter]{revtex4-2}
\usepackage{graphicx,bm,times}
\usepackage{amsmath}
\usepackage{amsfonts}
\usepackage{amssymb}
\usepackage{color}
\usepackage{hyperref}
\hypersetup{
	colorlinks = true,
	allcolors = {blue}
}

\usepackage[utf8]{inputenc}
\usepackage[T1]{fontenc}
\usepackage{amsmath,amssymb}
\usepackage{lineno}
\usepackage{float}
\usepackage{siunitx}

\mathchardef\mhyphen="2D

\begin{document}

\title{Novel electronic structures from anomalous stackings in NbS\textsubscript{2} and MoS\textsubscript{2}}

\author{Matthew D. Watson}
\email{matthew.watson@diamond.ac.uk}
\affiliation {Diamond Light Source Ltd, Harwell Science and Innovation Campus, Didcot, OX11 0DE, UK}

\author{Mihir Date}
\affiliation {Diamond Light Source Ltd, Harwell Science and Innovation Campus, Didcot, OX11 0DE, UK}
\affiliation {Max Planck Institute of Microstructure Physics, Weinberg 2, 06120 Halle (Saale), Germany}

\author{Alex Louat}
\affiliation {Diamond Light Source Ltd, Harwell Science and Innovation Campus, Didcot, OX11 0DE, UK}

\author{Niels B. M. Schr\"{o}ter}
\affiliation {Max Planck Institute of Microstructure Physics, Weinberg 2, 06120 Halle (Saale), Germany}

\date{\today}

\begin{abstract}
We show that in some transition metal dichalcogenides, minority regions of the cleaved sample surfaces show - unexpectedly and anomalously - a finite number of 2D electronic states instead of the expected 3D valence bands. In the case of NbS\textsubscript{2}, in addition to the typical spectrum associated with bulk 2Ha stacking, we also find minority regions with electronic structures consistent with few-layers of 3R stacking. In MoS\textsubscript{2} we find areas of both bulk 2Hc and 3R stackings, and regions exhibiting finite-layer quantisation of both types. We further find evidence for a more exotic 4Ha stacking of MoS\textsubscript{2}, in which the valence band maximum is quasi-2D. The results highlight how variation of the interlayer stacking of van der Waals materials beyond the commonly-reported bulk polytypes can yield novel electronic structures. 
\end{abstract}

\maketitle

\section{Introduction}

The great potential of transition metal dichalcogenides (TMDCs) for novel devices is from leveraging their thickness- and stacking-dependent electronic structures. The foremost experimental technique to reveal these is angle-resolved photoemission spectroscopy (ARPES), which has been frequently used for TMDCs in the form of 2D films, flakes, heterostructures, and cleaved bulk single crystals \cite{bussolotti_band_2023}. In the latter case, ARPES reveals both sharp spectral peaks from bands that disperse primarily in-plane, and broader, continuous, and photon energy-dependent spectral weight from bands with more 3D character \cite{strocov_three-dimensional_2012,el_youbi_fermiology_2021,watson_orbital-_2019}. Such a dichotomy is exemplified in 2H-(Mo,W)(S,Se)\textsubscript{2}, in which sharp valence band dispersions are observed around the $\rm\bar{K}$ point from quasi-2D states, but broad, continuous spectral weight found around $\bar{\Gamma}$, reflecting the $k_z$-dispersion of valence bands with 3D character \cite{riley_direct_2014,kim_determination_2016}. The valence bands with more 3D character are usually derived from orbitals with an out-of-plane character (e.g. $p_z$, $d_{3z^2-r^2})$ and are the most sensitive to variation of the interlayer stacking configuration, or "polytype" \cite{katzke_phase_2004,wolpert_polytypism_2023-1}. Moreover, in few-layer samples, it is these states which show the most obvious discretization, showing instead a finite number of bands depending on the number of layers \cite{jin_direct_2013, wilson_determination_2017, ulstrup_nanoscale_2019}.  

\begin{figure*}[t]
	\centering
	\includegraphics[width=0.95\linewidth]{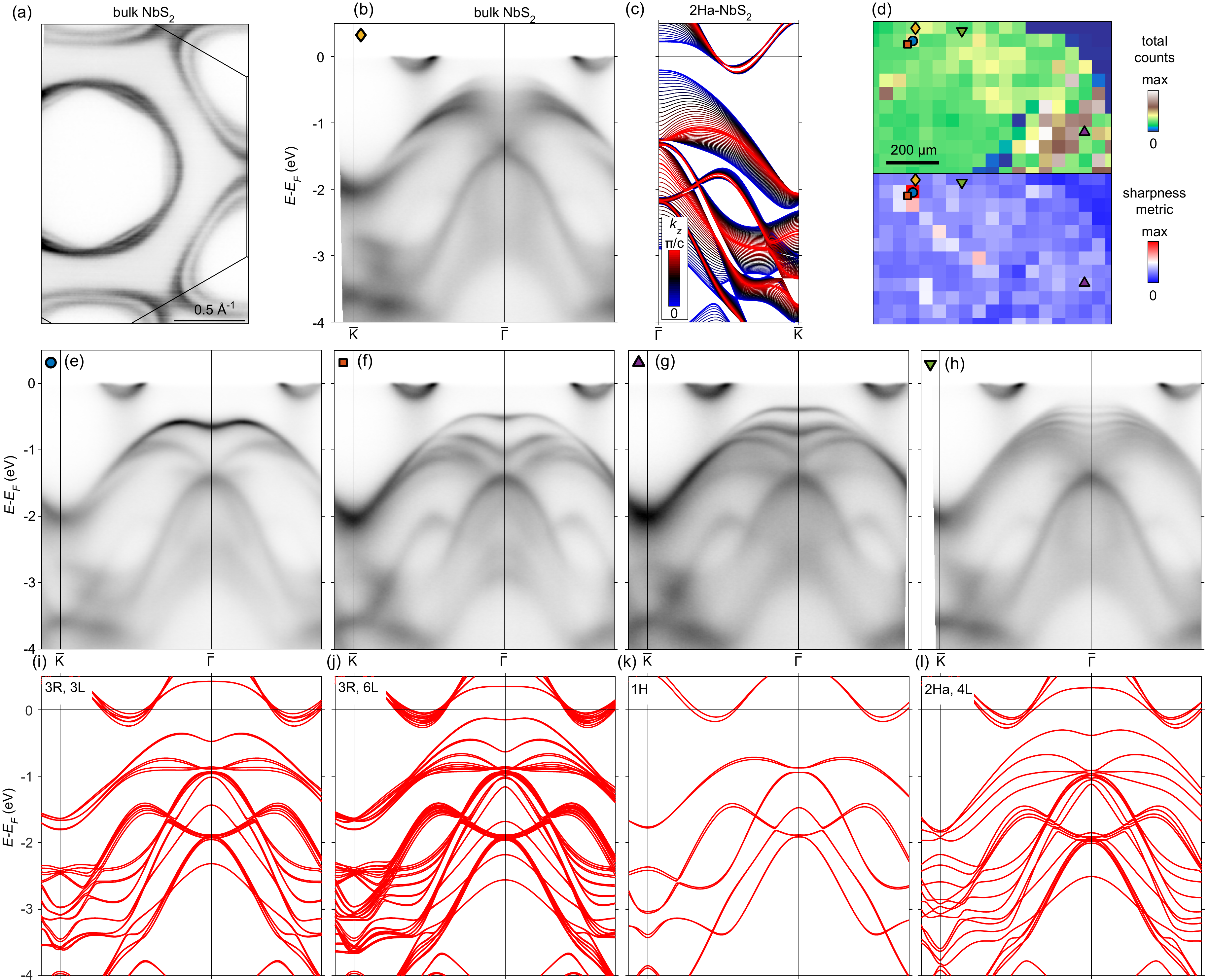}
	\caption{ (a) Fermi surface of bulk NbS\textsubscript{2}. (b) Dispersion in the $\mathrm{\bar{K}\mhyphen{}\bar{\Gamma{}}\mhyphen{}\bar{K}}$ direction measured on a bulk-like region. (c) DFT calculation of bulk NbS\textsubscript{2}, plotting dispersions at different $k_z$ according to the colormap. (d) Spatial map (total counts over an energy window of $\sim$ 1 eV below $E_F$) over a part of the cleaved sample. Lower panel plots the "sharpness" metric, defined in the supplemental material, which in this case indicates possible quantisation of the valence bands. (e-h) Dispersion measurements equivalent to (b), but measured on anomalous regions. The location of each within the spatial map is indicated by the coloured shapes. (i-l) Band dispersions calculated by DFT for (i) 3 layers in 3R stacking; (j) 6 layers of 3R stacking; (k) 4 layers in 2Ha stacking; (l) a single H layer. The quantized data in (e-g) are most consistent with few-layer 3R stackings as in (i,j). Both the measurements and calculations correspond to a cropped dispersion along M-K-$\Gamma$-K.}
	\label{fig1}
\end{figure*}

Here we use ARPES with a micro-focused beam spot ($\mu$ARPES) to scan the cleaved surface of nominally "2H" single crystal TMDCs and observe, remarkably, valence bands bearing signatures of finite-layer quantisation in some areas, as well as evidence for alternative polytypes. In NbS\textsubscript{2}, while much of the sample yields bulk-like spectra, the focused beam allows us to find minority regions with a rich variety of quantized valence band structures, although the quasi-2D bands at $E_F$ remain essentially unchanged. By comparison to density functional theory (DFT) calculations, we propose that these anomalous regions correspond to few-layer regions of NbS\textsubscript{2} with 3R-like stacking, rather than the normal 2Ha polytype. In the case of MoS\textsubscript{2} the normal stacking is 2Hc, but we identify one of the atypical areas as a region of periodic 3R stacking, and also find spectra containing quantisation signatures. We further find another region in which, unlike the 2H and 3R phases, the valence band maximum is quasi-2D, yet other bands at higher binding energies have 3D character. This indicates a bulk-like but unusual stacking structure, and we show the data is closely matched with calculations for a more exotic 4Ha polytype of MoS\textsubscript{2}. 

\section{Methods}
Single crystals of NbS\textsubscript{2} and MoS\textsubscript{2} were obtained commercially from HQ Graphene, Groningen. A key enabling technology for this study is capillary mirror-based synchrotron $\mu$ARPES \cite{koch_nano_2018} at the I05 beamline at Diamond Light Source, UK \cite{hoesch_facility_2017}, achieving a beam spot with approximate diameter (FWHM) of 4 $\mu{}$m. This enables fast scanning over large sample areas, while keeping the sample in the optical focus, and obtaining in a fraction a second a spectrum of sufficient quality and statistics to judge if a given location may show quantisation or other anomalous effects. Furthermore, the capillary mirror allows the user to routinely perform photon energy dependent scans while the beam spot on the sample remains absolutely fixed in both size and position, which is leveraged here to distinguish between 2D and 3D states. The combined energy resolution was 30-40 meV, and the beam polarisation was linear horizontal. The temperature was $\sim$30 K for the metallic NbS\textsubscript{2} and $\sim$100 K for the semiconducting MoS\textsubscript{2}, to mitigate charging effects. DFT calculations were performed using both VASP \cite{vasp3, vasp4, blochl_paw, paw2, pbe, Grimme-d3, pyprocar1} and Wien2k \cite{blaha_wien2k_2020, koller_improving_2012}, as detailed in the supplemental material (SM).

\section{Results}
\subsection{NbS\textsubscript{2}: patches of valence band quantisation}
2Ha-NbS\textsubscript{2} is a metallic TMDC, considered to be on the verge of a CDW instability, and a multiband two-gap superconductor with $T_c$ of $\approx$6~K \cite{guillamon_superconducting_2008}. Its fascinating phenomenology derives from the moderately correlated quasi-2D Nb $3d$-derived bands near E$_F$ \cite{el_youbi_fermiology_2021}, as shown in the Fermi surface in Fig.~\ref{fig1}(a) and the dispersion in Fig.~\ref{fig1}(b). However DFT calculations shown in Fig.~\ref{fig1}(c) predict that several of the deeper-lying valence bands should have substantial variation with $k_z$, particularly the uppermost fully-occupied band which has dominantly S $p_z$ character \cite{el_youbi_fermiology_2021,huang_probing_2022-1}. Thus, when taking into account the intrinsic $k_z$ broadening in ARPES \cite{strocov_intrinsic_2003}, one would expect to find a broad continuum of spectral weight at $\bar{\Gamma}$, in the range -1.5 to -0.5 eV, as indeed found in Fig.~\ref{fig1}(b), which is a spectrum representative of most of the sample.

Systematic scanning of the sample by $\mu$ARPES reveals, surprisingly, several regions exhibiting qualitatively differing spectra. While not obvious in the spatial map of the total counts in the analyser, the "sharpness" metric shown in Fig.~\ref{fig1}(d) - a measure proportional to the sum of gradients of the ARPES image measured at each spatial pixel, as further defined in the  Supplemental Material (SM) - highlights regions where the valence band dispersions are especially sharp. After optimising the alignment in these candidate regions, we obtained the anomalous spectra shown in \ref{fig1}(e-h). Approximately 7$\%$ of the sample surface showed some kind of anomalous spectrum, with spectra similar to Fig.~\ref{fig1}(f,g) being the most frequently observed.  

These spectra show remarkable qualitative differences from previous literature \cite{el_youbi_fermiology_2021,huang_probing_2022-1}, the bulk measurements in Fig.~\ref{fig1}(b), and each other. In place of the broad continuum of spectral weight at $\bar{\Gamma}$, as found in Fig.~\ref{fig1}(c), there emerge a discrete number of bands in the range -1.5 to -0.5 eV. The fact that these states are 2D in nature is already suggested by their sharp appearance, and further confirmed by photon energy dependence indicating the absence of any measurable out-of-plane dispersion (see SM). Nevertheless, the bands near $E_F$ are essentially unchanged, eliminating any proposal of 1T inclusions \cite{leroux_traces_2018}. These spectra cannot be explained by the reported bulk stacking configurations 2Ha and 3R \cite{el_youbi_fermiology_2021}, nor any alternative stacking \cite{katzke_phase_2004} with long-range periodicity in the out-of-plane direction.  

\begin{figure*}
	\centering
	\includegraphics[width=1\linewidth]{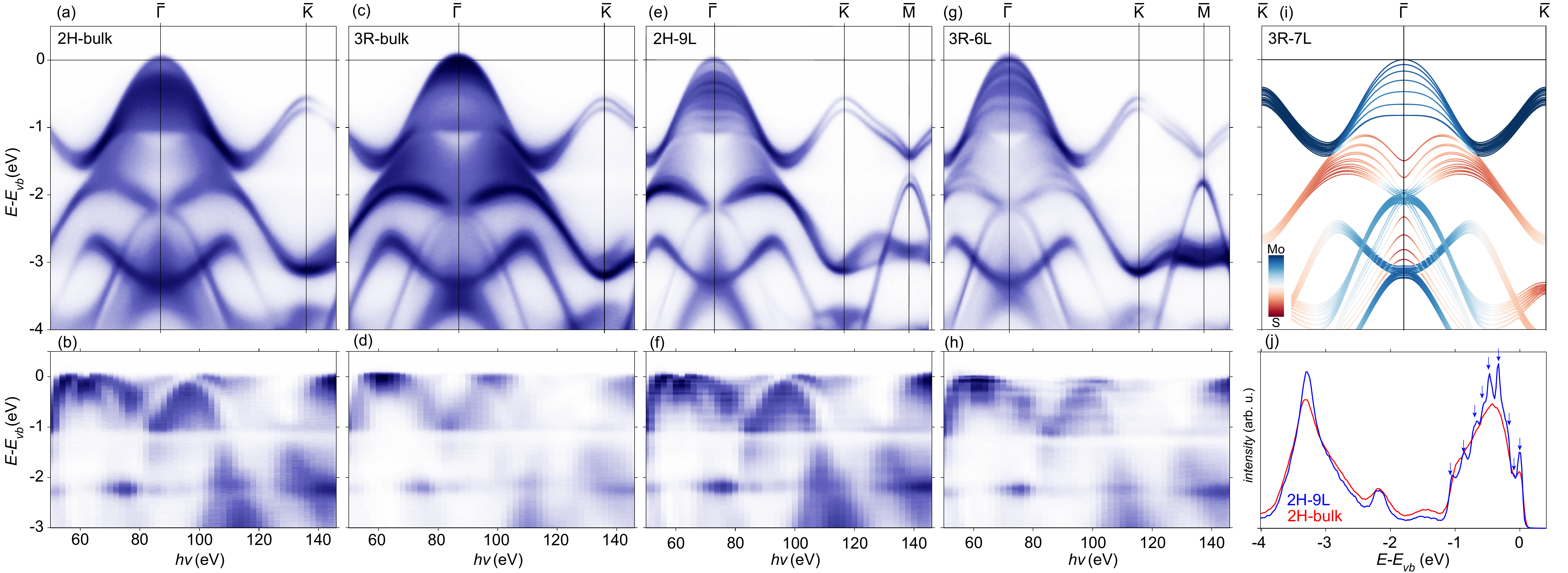}
	\caption{(a) Band dispersion in the $\rm \bar{K}\mhyphen\bar{\Gamma}\mhyphen\bar{K}\mhyphen\bar{M}$ direction, measured at $\mathrm{h\nu}$~=~100~eV and (b) photon energy dependence of the energy dispersion curve (EDC) at the $\Gamma$ point, both for a typical bulk region with 2Hc stacking. (c,d) Equivalent measurements on an area with 3R stacking, (e,f) a 9-layer system in the 2Hc stacking, a (g,h) a 6-layer system in the 3R stacking. (i) DFT calculation for a 7-layer stack of 3R-MoS\textsubscript{2}. (j) Comparisons of EDCs at $\bar{\Gamma}$ for 2Hc-bulk and 2Hc-9L (i.e. panels (a) and (e)), highlighting the spectral signatures of 9-fold quantisation.}
	\label{fig2}
\end{figure*}

Instead, the quantisation apparent in the data points towards near-surface stacking arrangements that lack long-range periodicity along the $c$ axis. The most naive scenario would be mechanical isolation of a few-layer system away from the bulk single crystal - a crack in the crystal a few layers below the surface, either present in the as-grown crystal, or possibly generated by the force of the cleave. Some larger-than-normal layer spacings (planar defects) inside NbS\textsubscript{2} crystals have indeed been observed by scanning transmission electron microscopy \cite{martino_unidirectional_2021}. In that case, we would expect the spectra to resemble calculations for freestanding few-layer systems. Alternatively, it may be that there is no mechanical exfoliation in these regions, but rather that the nominal bulk stacking is broken by at least one stacking fault, disrupting the periodicity along the $c$-axis: either a single fault near the surface, a boundary between regions of 2Ha and 3R stacking, or some more complex combinations of stackings without periodic repetition.  

For the few-layer cases, we performed \textit{ab initio} calculations on plausible stacking structures to compare with the experiments. We find that there is a good resemblance between the data showing quantisation and freestanding few-layer slabs in the 3R stacking: the best match to Fig.~\ref{fig1}(e) is a calculation for 3 layers in the 3R stacking in Fig.~\ref{fig1}(i), while Fig.~\ref{fig1}(f) is well-modelled by 6 layers of 3R as in Fig.~\ref{fig1}(j). While limited agreement can be found between Fig.~\ref{fig1}(e) and the 2Ha-stacked quad-layers (Fig.~\ref{fig1}(l)), the agreement is less good than for the 3R (Fig.~\ref{fig1}(i)). However we note that none of the measured spectra match with calculated valence band dispersions for freestanding monolayers (Fig.~\ref{fig1}(k)). 

As the number of layers increases, the number of possible structures scales rapidly with the number of layers ($\sim{}4^N$ if considering only trigonal prismatic coordination \cite{wolpert_polytypism_2023-1}), and the task of ascribing a real space stacking structure purely from electronic structure measurements is daunting. Fig.~\ref{fig1}(g) would be well-matched with a 9-layer stack in the 3R configuration (calculation not shown), but it would be difficult to exclude other possibilities, while for data such as Fig.~\ref{fig1}(h) where there are a larger number of quantized bands, implying many layers, there are so many possible structures to consider that determining the real space structure with precision and certainty is beyond reasonable expectation. 


To reconcile all the data on NbS\textsubscript{2} in Fig.~\ref{fig1}, we propose that the majority phase in our samples is 2Ha as is commonly reported to be the thermodynamically-preferred stacking for stoichiometric NbS\textsubscript{2} \cite{fisher_stoichiometry_1980}, but with inclusions of strata of 3R stacking, a few to several layers thick, present at the few-percent level. Sample cleavage exposes some such areas, perhaps preferentially, yielding the observed novel quantized band structures in select regions. 

\subsection{MoS\textsubscript{2}: 2H, 3R and quantisation}
We now consider "2H" MoS\textsubscript{2}, the semiconducting band structure of which is widely known. ARPES measurements on both thin film and exfoliated samples of MoS\textsubscript{2} \cite{jin_direct_2013,jung_spectral_2020,lee_time-resolved_2021,reidy_direct_2023} and MoSe\textsubscript{2} \cite{zhang_direct_2014} confirmed a direct band gap at the K point in a monolayer, but in bilayer or thicker samples the valence band maximum moves to the $\Gamma$ point. Few-layer samples exhibit a ladder of quantized energy levels at $\Gamma$, similar to quantum well states, that in asymptotic limit of infinite thickness converge to the bulk $k_z$ dispersion \cite{kim_determination_2016}.

In the mineralogy literature, it is well known that MoS\textsubscript{2} crystals, nominally preferring the 2Hc stacking for perfect stoichiometry, in fact host frequent stacking faults \cite{yang_periodic_2022}, and often inclusions of 3R phase \cite{strachan_3r-mos2_2021} the prevalence of which depend sensitively on impurity concentrations \cite{wickman_molybdenite_1970,newberry_polytypism_1979}. A variety of both periodic and non-periodic stackings have been observed by electron microscopy \cite{yang_periodic_2022}. Moreover, from a DFT perspective, the differences in formation energies between different stackings are known to be extremely low \cite{he_stacking_2014}. Indeed, MoS\textsubscript{2} is used industrially as a lubricant because of the low energy cost of sliding one layer with respect to adjacent layers. 

Upon scanning samples of MoS\textsubscript{2}, we again usually measure spectra similar to  Fig.~\ref{fig2}(a,b), ascribed to the normal 2Hc stacking of MoS$_2$, and closely resembling measurements in the literature \cite{kim_determination_2016}. However, we also encountered spectra with qualitative differences, such as the data presented in Fig.~\ref{fig2}(c,d). Here the spectral weight around $\bar{\Gamma}$ is also broad and continuous, indicating bulk electronic structure is being probed, but two specific spectral features identify this region as a bulk 3R stacking. First, the photon energy dependence varies more slowly in the 3R phase in Fig.~\ref{fig2}(d) compared with the 2Hc phase in Fig.~\ref{fig2}(b) which has more and different structure (especially apparent around $h\nu$=80 eV). This reflects the doubled periodicity along the $k_z$ direction in the 3R structure \cite{suzuki_valley-dependent_2014}, although we do not perform the mapping from $h\nu$ to $k_z$ coordinates here. Second, the 3R phase shows a slightly smaller splitting at the $\rm \bar{K}$ point of 0.14 eV, compared to 0.16 eV for the usual 2Hc; a direct comparison justifying this statement is plotted in the SM. This slightly smaller splitting is in agreement with an earlier literature report \cite{suzuki_valley-dependent_2014} and due to the absence of a bilayer splitting term, which adds in quadrature with the spin-orbit coupling in the 2H case, but is not present in the 3R structure. 

\begin{figure}
	\centering
	\includegraphics[width=1.0\linewidth]{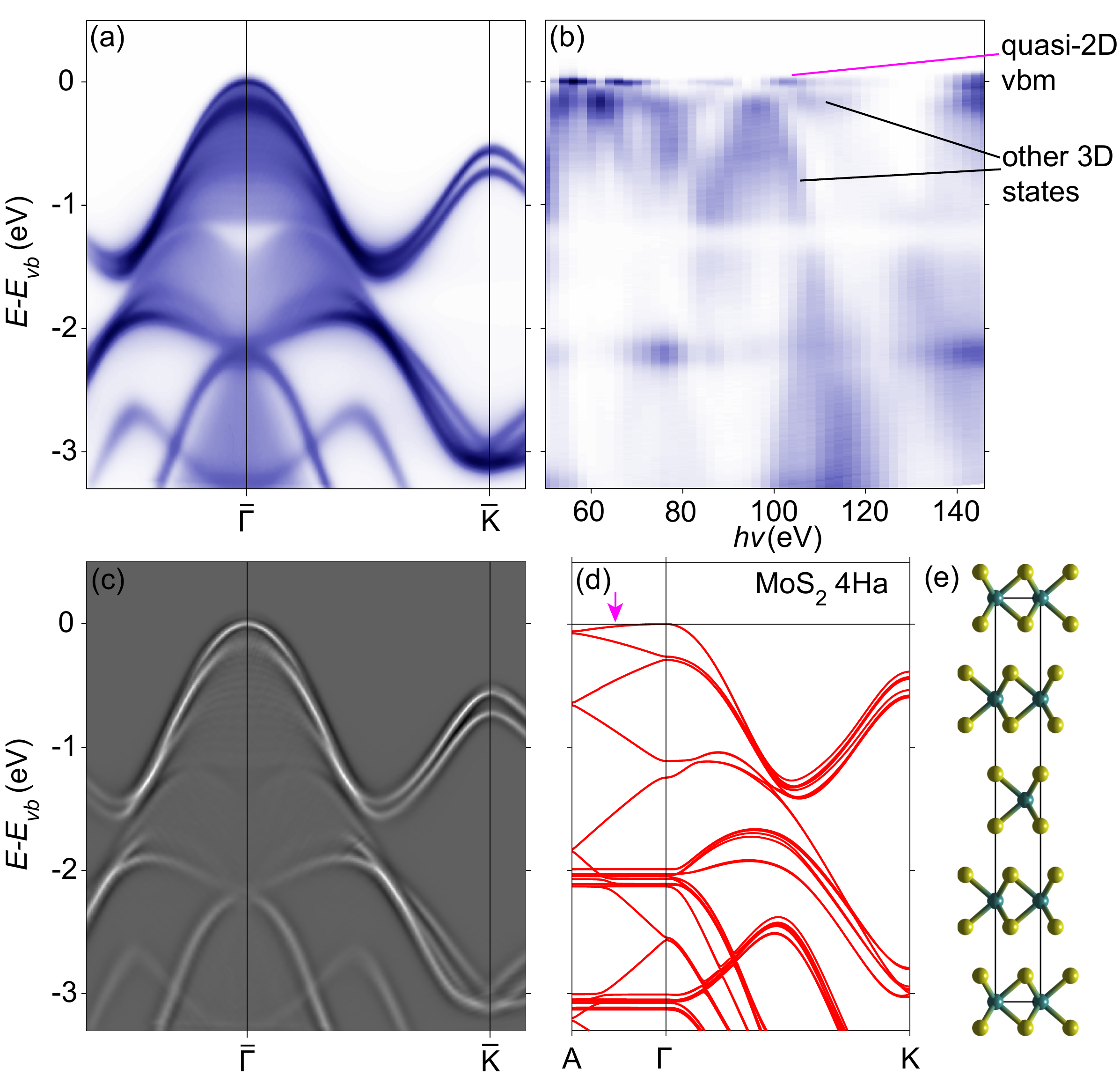}
\caption{Region exibiting quasi-2D valence band maximum, ascribed as 4Ha. (a) Sum of spectra as a function of photon energy (50-150~eV), showing a sharp valence band maximum. (b) dispersion at $\bar{\Gamma}$ with photon energy, highlighting the quasi 2D nature of the valence band maximum. (c) 2D curvature plot of (a). (d) Calculated band structure of MoS\textsubscript{2} in the 4Ha stacking, highlighting an almost 2D state with only ~60 meV dispersion along $\rm{}\Gamma-A$, and showing close resemblance with the data. (e) The unit cell of the 4Ha structure.}
	\label{fig3}
\end{figure}

Beyond the known bulk 2Hc and 3R stackings of MoS\textsubscript{2}, however, we find further spatial variation; analysis of a typical spatial map on MoS$_2$ presented in the SM reveals substantial variation on the length scales of 10-20 $\mu$m. In Fig.~\ref{fig2}(e,f) and Fig.~\ref{fig2}(g,h) we present spectra that resemble discretized versions of the usual bulk dispersion, with 9 and 6 states of quantisation of the valence band at $\bar{\Gamma}$ respectively, implying 9- and 6-layer structures (9L,6L). Interestingly, the photon energy dependence of the 6L in Fig.~\ref{fig2}(h) mirrors the photon energy dependence of the 3R phase (Fig.~\ref{fig2}(d)), while the 9L in Fig.~\ref{fig2}(f) follows the 2Hc spectral distribution (i.e. Fig.~\ref{fig2}(b)), except for the additional quantisation. Further, the overall lineshapes of the EDCs at $\mathrm{\bar{\Gamma}}$ shown in Fig.~\ref{fig3}(j) are very similar for the 2H and the 9L structure, except for the highlighted quantisation effects. Meanwhile, a calculation for a freestanding 7-layer structure in the 3R stacking in Fig.~\ref{fig2}(i) shows reasonable agreement with the data in Fig.~\ref{fig2}(g). We therefore ascribe these regions as 2Hc-9L and 3R-6L. 

\subsection{MoS\textsubscript{2}: evidence for 4Ha stacking}
Another kind of region merits particular attention, since its electronic structure qualitatively deviates from either the 2Hc- or 3R-MoS\textsubscript{2}. In these regions the top of the valence band is found to be unusually sharp, as shown in Fig.~\ref{fig3}(a), and most interestingly, the photon energy dependence in Fig.~\ref{fig3}(b) reveals it to be a quasi-2D state, without significant dispersion with $hv$. Moreover, unlike the quantized valence bands shown in \ref{fig2}(e-h), here this 2D state coexists with other valence bands that do vary with $hv$ and therefore have 3D character. We emphasise that energy of this state, found at the top of the valence band, is quite unlike the case for monolayer MoS\textsubscript{2} \cite{jin_direct_2013,jung_spectral_2020,lee_time-resolved_2021,reidy_direct_2023}, where the valence band at $\mathrm{\bar{\Gamma}}$ is at lower energy than those at $\mathrm{\bar{K}}$ - eliminating the possibility of a superposition of monolayer- and bulk-like spectra. Instead, to reconcile the observation of a single quasi-2D valence band within a bulk-like structure, we propose that this part of the sample may adopt the 4Ha stacking shown in Fig.~\ref{fig3}(e), previously reported for TaS\textsubscript{2} \cite{katzke_phase_2004}, but not previously considered for MoS\textsubscript{2}. Bulk calculations in this phase, shown in Fig.~\ref{fig3}(d), yield a valence band maximum that is nearly 2D with minimal dispersion along $\mathrm{\Gamma}-A$, and overall resembles the data very well, including fine features seen best in the curvature analysis in Fig.~\ref{fig3}(c). We therefore consider this to be strong evidence for the existence of minority regions of 4Ha-MoS\textsubscript{2} within the sample. 

\subsection{Scanning the family of 2H TMDCs}
\begin{figure}
	\centering
	\includegraphics[width=\linewidth]{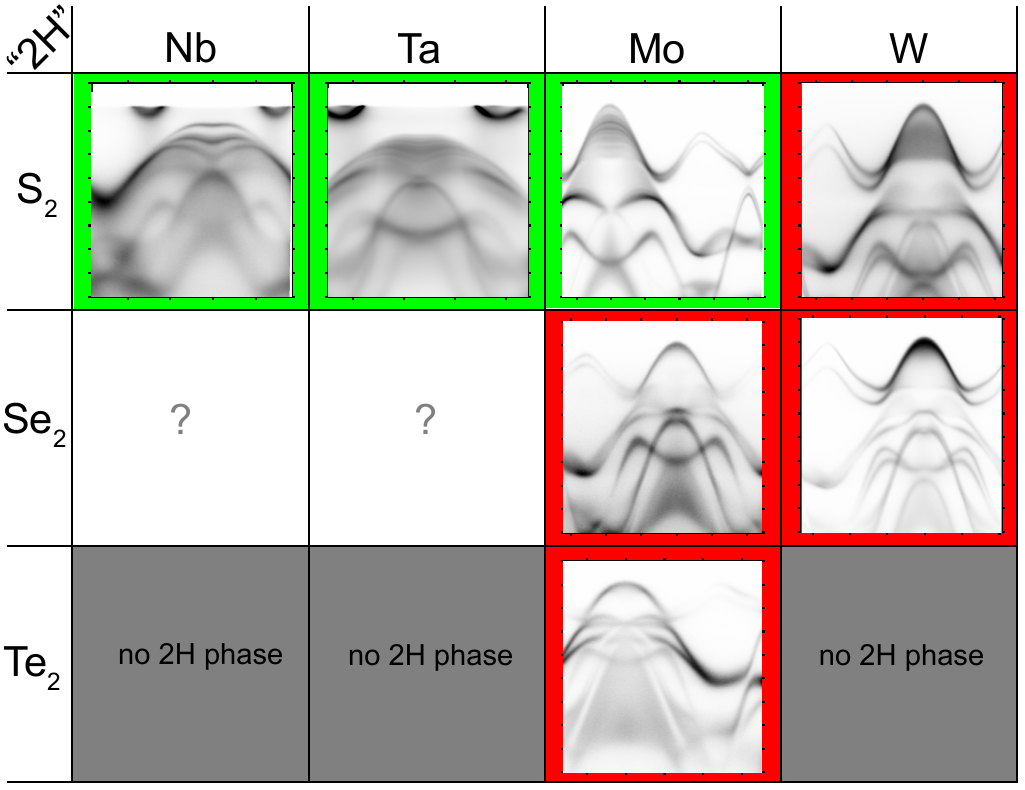}
	\caption{Summary of the transition metal dichalcogenides reported to stabilise in the 2H phase. Green and red borders indicate the presence or absence of valence band quantisation effects, with representative spectra shown. Tick marks represent 0.5 eV and 0.5 \AA{}$^{-1}$. NbSe\textsubscript{2} and TaSe\textsubscript{2} were not measured as part of this study.}
	\label{fig4}
\end{figure}

It is important to understand whether such anomalous stackings can be found ubiquitously in TMDCs, or only in the NbS\textsubscript{2} and MoS\textsubscript{2} discussed so far. We therefore performed $\mu$ARPES on most other members of the TMDCs with nominal "2H" stackings, summarised in Fig.~\ref{fig4}. In the case of the other semiconducting 2H materials, MoSe\textsubscript{2}, WSe\textsubscript{2}, WS\textsubscript{2}, and MoTe\textsubscript{2}, we performed spatial mapping with $\mu$ARPES on cleaves of each material, but only ever found the same 3D band structure. The case of metallic TaS\textsubscript{2} is very interesting, since the Ta compounds have the highest number of reported polytypes \cite{katzke_phase_2004}. TaS\textsubscript{2} has many complexities including charge density wave order that go beyond the scope of this study, but similar to the NbS\textsubscript{2} we observed finite-layer quantisation effects in a few regions. 

\section{Discussion}
The observation of anomalous stackings within our results is limited to the "2H" sulfides, excluding WS\textsubscript{2}. We speculate that there may be a link between these observations and the physical properties of the samples: compared to the selenides, the sulfides are typically much thinner and flakier, while selenide crystals are often thicker, more rigid, and often show clear crystal facets. However it is likely to be the case that different sample growth and preparation methods might promote or discourage the formation of anomalous stackings. Therefore our results do not exclude that anomalous stackings could be found in other studies of selenides (we note that some anomalous quantisation was previously reported in TaSe\textsubscript{2} \cite{Li_quantization_2021}), nor do our results necessarily imply that minority phases or quantised bands will be found in all samples of MoS\textsubscript{2} or NbS\textsubscript{2}. We consistently found spectra similar to Fig.~\ref{fig1}(e-h) on the NbS\textsubscript{2} on at least 3 different cleaved samples, albeit all from the same growth batch. On the MoS\textsubscript{2} we consistently found some spatial variation and 3R regions, however the clear many-layer quantisation effects in Fig.~\ref{fig2}(e-h) were found only on one cleave. However regions resembling Fig.~\ref{fig3}, attributed as 4Ha stacking, were encountered several times.   

 As $\mu$ARPES becomes increasingly popular, we anticipate that further discoveries of anomalous stackings will follow, even in systems which have been previously well-studied, such as the case of MoS\textsubscript{2} reported here. However, a shortcoming of our approach is the lack of correlative local structural information on the near-surface stacking. With only electronic structure information, we are limited to making assignments based on comparison with electronic structure calculations. 
 
 Nevertheless, the results demonstrate that $\mu$ARPES can be a route to measuring electronic structures of minority phases that may naturally occur within as-grown crystals. The ascribed 4Ha stacking of MoS\textsubscript{2} in Fig.~\ref{fig3} exemplifies the potential for new discovery with this approach. The presence of a 2D rather than 3D valence band maximum in this polytype would be expected to radically change its optical and electronic properties, such as the mobility of the hole-like carriers. The finding of 4Ha-MoS\textsubscript{2}, at least as a minority phase, should therefore encourage both theoretical and synthesis investigations into other possible exotic stacking possibilities of this family of technologically-relevant materials. 

\section{Conclusion}
In conclusion, we have characterised novel electronic structures from minority polytypes and near-surface stacking faults in NbS\textsubscript{2} and MoS\textsubscript{2}. By comparison with DFT calculations, we suggest that several-layer strata of 3R-NbS\textsubscript{2} are present in the sample of nominal 2Ha composition, which are exposed upon cleavage, yielding the quantized 2D valence bands. In the case of MoS\textsubscript{2}, as well as the typical 2Hc spectra we find bulk-like 3R regions, a previously unknown stacking which we associate with the 4Ha polytype, and further spectra with quantisation effects characteristic of multilayers. Going beyond these two prototypical TMDCs, we find analagous effects in TaS\textsubscript{2}, but did not find any anomalous regions in Mo(Se,Te)\textsubscript{2} or W(S,Se)\textsubscript{2}. The discovery of these quantized features, alongside the quality of the data observed on both the bulk-like and quantized regions, highlight the benefits of performing $\mu$ARPES even on samples thought to be "homogeneous single crystals" and demonstrates how varying the $c$-axis stacking of 2D materials can yield novel electronic structures.

\begin{acknowledgments}
We thank T. K. Kim, C. Cacho, D Biswas, and N. Wilson for insightful discussions. The MoTe\textsubscript{2} data in Fig.~\ref{fig4} were kindly shared by Jonas Krieger. We thank Shigemi Terakawa for experimental support. 
We acknowledge Diamond Light Source for time on beamline I05 under proposals NT31067, NT31222 and SI31407. 
\end{acknowledgments}

~\\ 

\section{References}
\bibliographystyle{apsrev4-2}


%

\end{document}